\documentclass[twocolumn,amstext,amsmath,amssymb,prc,superscriptaddress,floatfix,showpacs]{revtex4-1}
 
\usepackage{slashed}
\usepackage{lineno}
\usepackage{graphicx}
\usepackage{color}
\usepackage{cancel}
\usepackage{ulem}
\usepackage{subfigure}
\usepackage{mathrsfs}
\usepackage{dcolumn}
\usepackage{xspace}
\usepackage{bbold}
\usepackage{hyperref}
\newcommand{\dslash}{\ensuremath{\partial\hspace{-1.2ex} /}}

\newcommand{\ua}{\ensuremath{U(1)_A}}

\def\Eq#1{Eq.~(\ref{#1})}
\def\Fig#1{Fig.~\ref{#1}}
\def\App#1{App.~\ref{#1}}
\def\Tab#1{Tab.~\ref{#1}}
\def\Eqs#1{Eqs.~(\ref{#1})}

\def\roughly#1{\mathrel{\raise.3ex\hbox{$#1$\kern-.75em%
\lower1ex\hbox{$\sim$}}}}

\newcommand{\Omegaqq}{\ensuremath{\Omega_{\bar{q}q}}}

\newcommand{\vev}[1]{\ensuremath{\left\langle #1 \right\rangle}}

\newcommand{\diag}{\ensuremath{\operatorname{diag}}}

\def\ma0{m_{a_{0}}}
\def\mf0{m_{f_{0}}}

\newcommand{\Sigmamin}{{\bar{\Sigma}}}
\newcommand{\sigx}{\sigma_x}
\newcommand{\sigy}{\sigma_y}

\begin{document}

\title{Fluctuations and the axial anomaly with three quark flavors} \author{Mario Mitter}
\email[E-Mail:]{m.mitter@thphys.uni-heidelberg.de}
\affiliation{Institut f\"{u}r Theoretische Physik,
  Ruprecht-Karls-Universit\"{a}t Heidelberg, D-69120 Heidelberg,
  Germany\\[0.5ex]} \affiliation{Institut f\"{u}r Theoretische Physik,
  Goethe-Universit\"{a}t Frankfurt, D-60438 Frankfurt,
  Germany\\[0.5ex]} \author{Bernd-Jochen Schaefer}
\email[E-Mail:]{bernd-jochen.schaefer@theo.physik.uni-giessen.de}
\affiliation{Institut f\"{u}r Theoretische Physik,
  Justus-Liebig-Universit\"{a}t Gie\ss en, D-35392 Gie\ss en,
  Germany\\[0.5ex]} \affiliation{Institut f\"ur Physik,
  Karl-Franzens-Universit\"at Graz, A-8010 Graz, Austria}


\begin{abstract}
  The role of the axial anomaly in the chiral phase transition
  at finite temperature and quark chemical potential is investigated
  within a non-perturbative functional renormalization group
  approach. The flow equation for the grand potential is solved to
  leading-order in a derivative expansion of a three flavor
  quark-meson model truncation.  The results are compared with a
  standard and an extended mean-field analysis, which facilitates the
  exploration of the influence of bosonic and fermionic fluctuations,
  respectively, on the phase transition.  The influence of
  $\ua$-symmetry breaking on the chiral transition, the location of a
  possible critical endpoint in the phase diagram and the quark mass
  sensitivity is studied in detail.
\end{abstract}

\maketitle

\section{Introduction}
\label{sec:intro}

Quantum Chromodynamics (QCD) with $N_f$ flavors of massless quarks has
a global $U(N_f)_L \times U(N_f)_R$ symmetry which is spontaneously
broken in the low-energy hadronic sector of QCD by the formation of
non-vanishing chiral condensates.  The included axial $\ua$-symmetry
is violated by quantization yielding the axial or chiral anomaly
\cite{Adler:1969er, *Bell:1969ts, *Fujikawa1979b} which is related to
the $U(1)_A$-problem \cite{Weinberg:1975ui, Christos:1984tu,
  Hooft1986}.  The large mass of the $\eta'$-meson can be explained by
an instanton induced 't~Hooft determinant \cite{'tHooft:1976up,
  *'tHooft:1976fv} and is linked to the topological susceptibility of
the pure gauge sector of QCD \cite{Witten1979, *Veneziano1979}.

At high temperatures and baryon densities QCD predicts a transition
from ordinary hadronic matter to a chirally symmetric phase, whose
detailed symmetry restoration pattern is not yet fully clarified. At
baryon densities a few times of normal nuclear density and 
relatively low  temperatures 
 a color-flavor locked phase is expected to appear
whereas the situation at smaller intermediate densities is less clear
\cite{Alford:2007xm} and inhomogeneous phases might additionally arise \cite{Nickel:2009wj}.
A deeper understanding of the nature of the chiral phase
transition is not only important on the theoretical side, but also
plays a crucial role in relativistic heavy-ion
experiments~\cite{Friman:2011zz}.  On the one hand, it is
well-established that the $\ua$-symmetry is restored at sufficiently
high temperatures and chemical potentials \cite{Gross:1980br,
  *Schafer:1996wv, *Schafer:2002ty}.  
On the other hand, it is an open issue whether an effective $\ua$-symmetry
  restoration occurs at the chiral transition temperature for three
  physical quark masses. For related two and three flavor
  investigations in the chiral limit see e.g. \cite{Cohen:1996ng,
    *Lee:1996zy, *Birse:1996dx, *Dunne:2010gd}.

Recent analyses of experimental data by the PHENIX and STAR
collaborations at the Relativistic Heavy Ion Collider (RHIC) have
revealed a drop in the $\eta'$-meson mass at the chiral crossover
temperature \cite{Vertesi:2009wf, *Csorgo2010}. This observation is
interpreted as a sign of an effective $\ua$-symmetry restoration
already at the chiral transition temperature \cite{Kapusta:1995ww}.

Similar conclusions can be drawn from several recent lattice QCD
investigations \cite{Bazavov:2012qja, Cossu:2012gm,
  *Cossu:2013uua}. 
 Unfortunately, these investigations are still
hampered by a sign problem at finite chemical potential,
e.g.~\cite{Muroya:2003qs, *Philipsen:2007rj}.  Other non-perturbative
approaches without sign problem are based on continuum methods such as
the functional renormalization group (FRG) \cite{Berges:2000ew,
  *Aoki:2000wm, *Schaefer:2006sr, *Gies2006, *Braun:2011pp,
  *Pawlowski:2005xe}. Recently, functional methods have been
prosperously applied to the $U(1)_A$ problem, see
e.g. \cite{Pawlowski:1996ch, *vonSmekal:1997dq, *Alkofer:2008et, *Benic:2011fv}, 
the low-energy QCD sector at finite temperature and chemical potentials,
e.g. \cite{Braun:2009gm, *Pawlowski:2010ht, *Fischer:2013eca,
  *Fischer:2011mz, *Fischer:2012vc} as well
as QCD-like effective theories.  Usually, such effective
investigations are performed with two flavors, assuming a strong axial
anomaly and taking only the $(\sigma,\vec \pi)$-multiplet into account
\cite{Jungnickel1996b, Berges:1997eu, *Schaefer1999, *Braun:2003ii,
  Schaefer:2004en, *Schaefer:2006ds}, see \cite{Gies:2002hq, *Braun:2008pi} for a
more elaborate one flavor approach.  In view of recent lattice and
experimental observations the assumption of a strong axial anomaly
seems to be not justified, at least in the vicinity of the chiral
transition.  However, it has been shown in purely bosonic theories
that the chiral transition crucially depends on the fate of
$\ua$-symmetry violating operators at the chiral transition
\cite{Pisarski1984a, SchaffnerBielich:1999uj, Lenaghan:2000kr}.  As a consequence, the proper
implementation of the $\ua$-anomaly is also important in model
investigations at intermediate chemical potentials. In particular,
this might affect a possibly existing critical endpoint in the QCD
phase diagram \cite{Hatsuda:2006ps, *Yamamoto:2007ah, *Chandrasekharan:2007up, Chen2009}.

This work is an extension of previous analyses within effective linear
sigma models with two \cite{Schaefer:2004en, *Schaefer:2006ds} and
three quark flavors \cite{Schaefer:2008hk} to the more realistic case
where the three flavor dynamics on the chiral phase transition is
included beyond mean-field approximations. The restoration of the
chiral $SU(3)_L\times SU(3)_R$ as well as the axial $\ua$-symmetry
with temperature and quark chemical potential are addressed. The
breaking of the $\ua$-symmetry in the Lagrangian is implemented by an
effective Kobayashi-Maskawa-'t~Hooft determinant
\cite{Kobayashi:1970ji, 'tHooft:1976up, *'tHooft:1976fv} which models
the axial $\ua$-anomaly.

The outline of this work is as follows: In the next
Sec.~\ref{sec:frg_model} we examine a $U(N_f)\times U(N_f)$-symmetric
chiral quark-meson model with axial $\ua$-symmetry breaking. In order
to study the influence of thermal and quantum fluctuations on the
chiral phase transition including the axial anomaly various
approximations of the grand potential are considered. In
Sec.~\ref{sec:mf} we briefly summarize the mean-field approximation of
the grand potential of the three flavor model, where mesonic
fluctuations are ignored, c.f.~\cite{Schaefer:2008hk}. These
fluctuations are included with the functional renormalization group
method which we discuss in a leading-order derivative expansion of the
effective action in Sec.~\ref{sec:frg}.  Our numerical results on the
chiral phase transition, the dependency on the axial anomaly and the
quark mass sensitivity are collected in Sec.~\ref{sec:numres}.  We
conclude and summarize in Sec.~\ref{sec:summary}. Technical details of
the FRG implementation are given in the appendices.

\section{$\ua$-symmetry breaking in Chiral Models}
\label{sec:frg_model}

Quark-meson models often serve as an effective description of
low-energy QCD with $N_f$ quark flavors. They consist of a chirally
invariant linear sigma model, typically for (pseudo)scalar
mesonic degrees of freedom, $\Sigma$, a Yukawa-type quark-meson
vertex and a bilinear quark action \cite{Schwinger:1957em,
  *Gell-Mann:1960np}.

In general, the Euclidean Lagrangian of the mesonic sector with a
global chiral $U(N_f)_L\times U(N_f)_R$ flavor symmetry has the form
\cite{Meyer-Ortmanns:1992pj, *Meyer-Ortmanns:1994nt, Jungnickel1998,
  Lenaghan:2000ey, *Scavenius:2000qd, Patkos:2012ex}
\begin{eqnarray}\label{eq:meson}
  {\mathcal L}_\textrm{m} & = & \mathsf{tr}\!\left[ \partial_\mu \Sigma \partial_\mu
    \Sigma^\dag \right] +U\bigl(\left\{\rho_i\right\}\bigr)\ ,
\end{eqnarray}
where the potential $U$ is a function of chiral invariants $\rho_i$
defined by
\begin{eqnarray}\label{eq:rhoi}
  \rho_i & = & \mathsf{tr}\!\left[\left(\Sigma
      \Sigma^\dag\right)^i\right]\,,\quad  i =
    1,\ldots, N_f\ . 
\end{eqnarray}
It is also possible to construct higher chiral invariants with
$i>N_f$, but these invariants can be expressed in terms of the lower
ones with $i\leq N_f$ \cite{Jungnickel1996b}.  In addition, only the
invariants $\rho_1$ and $\rho_2$ correspond to renormalizable
interactions in four spacetime dimensions and $\rho_1$ is the only
invariant quadratic in the fields.

The ($N_f \times N_f$)-matrix field $\Sigma$ parametrizes the
scalar $\sigma_a$ and the pseudoscalar $\pi_a$ meson multiplets
\begin{eqnarray}\label{eq:sigma}
\Sigma & = & T^a\left(\sigma_a+ i \pi_a\right)\ , 
\end{eqnarray}
where the $N_f^2$ Hermitian generators of the $U(N_f)$ symmetry are
normalized by $\mathsf{tr}[T^aT^b]=\delta^{ab}/2$.  For three flavors the
generators are $T^a=\lambda^a/2$ with the standard Gell-Mann matrices
$\lambda^a$ and $\lambda^0 = \sqrt{2/3}$ $\mathbb{1}$.  Explicit
chiral symmetry breaking can be implemented by adding
\begin{eqnarray}\label{eq:explbreaking}
  \mathsf{tr}\!\left[C(\Sigma+\Sigma^\dag)\right]\ ,\quad C =
  T^ac_a\ ,
\end{eqnarray}
to the Lagrangian (\ref{eq:meson}) yielding non-vanishing Goldstone
boson masses.  Adjusting the constant parameters $c_a$, different
explicit symmetry breaking patterns are possible \cite{Jungnickel1998,
  *Lenaghan:2000ey, *Scavenius:2000qd}.

As mentioned in the introduction, the (axial) $\ua$-symmetry breaking
can be implemented in the effective Lagrangian on the tree-level by
adding the lowest dimensional $\ua$-symmetry violating operator, the
Kobayashi-Maskawa-'t~Hooft interaction term \cite{Kobayashi:1970ji,
  'tHooft:1976up, *'tHooft:1976fv},
\begin{eqnarray}\label{eq:xi}
 \xi & = & \det\left[\Sigma\right] + \det\left[\Sigma^\dag\right]\ ,
\end{eqnarray}
to the Lagrangian which represents a determinant in flavor space. On
the quark level, this interaction corresponds to a flavor- and
chirality-mixing $2N_f$-point-like interaction with $N_f$ incoming and
$N_f$ outgoing quarks.  From a phenomenological point of view this
term is important to properly describe, for example, the $\eta$ and
$\eta'$ meson mass splitting \cite{Hooft1986, Klabucar:2001gr, *Kovacs:2007sy, *Jiang:2012wm, *Kovacs:2013xca}.  
It breaks the axial
$\ua$-symmetry, but is invariant under the $SU(N_f)_L\times SU(N_f)_R$ and
$U(1)_V$-symmetry. However, other axial symmetry breaking terms are
possible, cf. e.g. ~\cite{Jungnickel1996b}. For example, a
$\ua$-symmetry breaking term proportional to ($\det\left[\Sigma\right] -
\det\left[\Sigma^\dag\right]$), would violate the discrete $\Sigma \to
\Sigma^\dag$ symmetry and is discarded due to the required $CP$
invariance of the model.  The $\ua$-symmetry breaking term via
\Eq{eq:xi} scales with the meson (quark) fields to the power $N_f$
($2N_f$), which leads to qualitative differences as the number of
flavors is increased \cite{Kobayashi:1970ji, Jungnickel1996b}.  In the
mesonic formulation, the $N_f=1$ case yields a linear explicit breaking
term, whereas for $N_f=2$ the determinant corresponds to a mesonic
mass term.  Important $N_f$-dependent effects on the chiral transition
are anticipated.

We focus on three quark flavors $q = (u, d, s)$ with an exact $SU(2)$
isospin symmetry in the light quark sector, i.e. two light flavors
are degenerate with $m_u = m_d \equiv m_l$. The renormalizable and
chirally symmetric potential with the $\ua$-anomaly is expanded as
\begin{equation}
  \label{eq:2}
  U(\rho_1,\rho_2,\xi) = m^2\rho_1 + \lambda_1\rho_1^2 + \lambda_2 \rho_2 - c\xi\ ,
\end{equation}
where we have introduced four parameters $m^2$, $\lambda_{1}$,
$\lambda_2$ and $c$ with values such that the potential is bounded
from below. The parameter $c$ represents the strength of the cubic
$\ua$-symmetry violating determinant and is temperature and density
dependent in general \cite{Chen2009}. In the instanton picture the
anomaly strength in the vacuum is proportional to the instanton
density and can be estimated perturbatively \cite{'tHooft:1976up,
  *'tHooft:1976fv}.  In total, without the
Kobayashi-Maskawa-'t~Hooft term, i.e. for $c=0$, the model has
an $U(1)_V\times SU(N_f)_L\times SU(N_f)_R\times U_A(1)$ symmetry
which reduces to an $U(1)_V\times SU(N_f)_L\times SU(N_f)_R$ symmetry
for non-vanishing $c$ apart from
multiple covering of the groups.  The potential is of order ${\mathcal
  O}(\Sigma^4)$ and all invariants obey the discrete symmetry $\Sigma
\to \Sigma^\dag$.

Finally, the quark-meson (QM) model is obtained by coupling quarks in
the fundamental representation of $SU(N_f)$ to the mesonic
sector which yields the Lagrangian ${\mathcal L}_\textrm{qm} = {\mathcal
  L}_\textrm{q} + {\mathcal L}_\textrm{m}$.  In the quark Lagrangian
\begin{eqnarray}\label{eq:model}
 {\mathcal L}_\textrm{q} & = & \bar{q} \left(\dslash  + \hat\mu\gamma^4 +
    h \Sigma_5 \right)q\ ,
\end{eqnarray}
a flavor and chirally invariant Yukawa interaction with strength $h$
of the mesons to the quarks has been introduced where
\begin{equation}\label{eq:sigma5}
\Sigma_5 =
T^a\left(\sigma_a+i\gamma_5\pi_a \right)\ .
\end{equation}
The quark chemical potential matrix $\hat\mu$ is diagonal in flavor
space $\hat\mu = \diag (\mu_l, \mu_l, \mu_s)$
 and we will consider in the
following only one flavor symmetric quark chemical potential
$\mu\equiv \mu_l = \mu_s $.

Appropriate order parameters for spontaneous chiral symmetry breaking
are the quark condensates that are related via bosonization to vacuum
expectation values of the corresponding scalar-isoscalar mesonic
fields. For three quark flavors the corresponding non-vanishing
condensates that carry the proper quantum numbers of the vacuum are
$\vev{\sigma_a}$ with $a=0,3,8$. For an exact $SU(2)$ isospin symmetry
$\vev{\sigma_3}$ vanishes and only the remaining two condensates are
independent.

As argued in \cite{Schaefer:2008hk} it is advantageous to rotate the
scalar singlet-octet $(0 - 8)$ basis into the non-strange--strange $(x
- y)$ basis by
\begin{eqnarray}\label{eq:sns_basis}
 \left(\begin{array}{c}\sigma_{x}\\ \sigma_y\end{array}\right)
& = &
\frac{1}{\sqrt{3}}\left(
\begin{array}{cc}
\sqrt{2} & 1\\
1 & -\sqrt{2}
\end{array}
\right)
\left(\begin{array}{c}\sigma_{0}\\ \sigma_8\end{array}\right)\ .
\end{eqnarray}
In this case, the explicit symmetry breaking term simplifies to
$\mathsf{tr}\!\left[C(\Sigma+\Sigma^\dag)\right] \to c_x\sigx+
c_y\sigy$ with the modified non-strange $c_x$ and strange $c_y$
explicit symmetry parameters. For the vacuum expectation value we
obtain $\vev\Sigma = \diag(\langle\sigma_x\rangle/2,
\langle\sigma_x\rangle/2,\langle\sigma_y\rangle/\sqrt{2})$.

\section{Mean-field Approximation}
\label{sec:mf}

We begin with a mean-field analysis of the three flavor model and
derive the grand potential.  In the mean-field approximation (MFA) of
the path integral for the grand potential, certain quantum and thermal
fluctuations are neglected.  In case of the quark-meson model the
mesonic quantum fields are usually replaced by constant classical
expectation values. Only the integration over the fermionic degrees of
freedom is performed, which additionally yields a divergent vacuum
contribution to the grand potential. In the standard MFA this vacuum
term is simply ignored. However, the quark-meson model is
renormalizable and the inclusion of the divergent vacuum contribution
to the grand potential is possible. The consideration of the vacuum
terms represents a first step beyond the standard MFA, see
\cite{Andersen:2011pr, Skokov:2010sf} for the influence of the vacuum
fluctuations on the thermodynamics and \cite{Chatterjee:2011jd,
  *Mao:2009aq, *Schaefer:2011ex, Strodthoff:2011tz} for corresponding
investigations in other models.  In the following we will employ the
standard (no-sea) MFA where the vacuum term in the grand potential is
omitted, in order to straightforwardly compare our results with
previous works.  Later on we will confront our MFA analysis with
various renormalization group results where the vacuum terms are
included.

In the no-sea MFA the resulting grand potential consists of two
contributions: the purely mesonic potential $V$ with a linear explicit
chiral symmetry breaking term and the quark/antiquark contribution
\Omegaqq
\begin{eqnarray} \label{eq:grandpot}
  {\Omega}_\text{MF}(T,\mu) & = &
  \Omegaqq (T,\mu;{\Sigmamin}) +V({\Sigmamin})\ ,
\end{eqnarray}
evaluated at the minimum $\Sigmamin \equiv \vev{\Sigma}$ of the
potential.  Explicitly, the mesonic potential is given by
\begin{eqnarray} 
  V(\Sigma)  & = & \frac{m^2}{2}(\sigma_x^2 + \sigma_y^2)
  -c_x\sigma_x-c_y\sigma_y - \frac{c}{2\sqrt{2}}\sigma_x^2\sigma_y\nonumber
  \\
  & + & \frac{2\lambda_1+\lambda_2}{8}\sigma_x^4 + \frac{\lambda_1 }{2}\sigma_x^2\sigma_y^2 
  + \frac{\lambda_1+\lambda_2}{4}\sigma_y^4   
\end{eqnarray}
and the quark/antiquark contribution without the vacuum term reads
\begin{eqnarray}
\Omegaqq & \!=\! & -{2N_cT} \sum\limits_{f=1}^{N_f}
  \int\!\!\!\frac{ d^3p}{(2\pi)^3}\bigg\{
  \ln\left[1+\text{e}^{-\left(E_{p,f}+
        \mu\right)/T}\right] +\nonumber \\
  & & \phantom{-{2N_cT} \sum\limits_{f=1}^{N_f}  \int }
\ln\left[1+\text{e}^{-\left(E_{p,f}-
        \mu\right)/T}\right]\bigg\}\  .
\end{eqnarray}
The quark/antiquark single-quasiparticle energies are defined by
$E_{p,f}^2 =\vec{p}^{\:2} + m_f^2$ with the field-dependent
eigenvalues $m_f^2$ of the mass matrix $h^2 \Sigma \Sigma^\dag$, which
yield the constituent quark masses, when evaluated at the minimum
$\Sigmamin$.  In the non-strange--strange ($x$-$y$) basis and for
$SU(2)$ isospin symmetry the masses simplify to
\begin{eqnarray}
  \label{eq:mq}
 m_{l}  =  h\frac{\sigma_x}{2}\ , & \  & m_{s}  =  h\frac{\sigma_y}{\sqrt{2}}\ ,
\end{eqnarray}
where the index $l$ labels the two degenerate light (up and down) flavors
and the index $s$ stands for the strange flavor.

Finally, the two order parameters for the non-strange and strange
chiral phase transition $\langle\sigx\rangle$ and
$\langle\sigy\rangle$ are obtained as the (global) minimum of the
grand potential (\ref{eq:grandpot}) and are functions of the
temperature and quark chemical potential, cf~\cite{Schaefer:2008hk}.

\section{Functional Renormalization Group Analysis}
\label{sec:frg}

For the non-perturbative analysis of the three flavor model we employ
a functional renormalization group (FRG) equation.  One possible
implementation of the Wilsonian renormalization group idea is based on
the effective average action approach pioneered by
Wetterich~\cite{Wetterich1993d}.  The scale evolution of the effective
average action $\Gamma_k[\Phi]$ with arbitrary field content $\Phi$
including fermionic and bosonic fields is governed by
\begin{eqnarray}\label{eq:WetterichRG}
  \partial_t\Gamma_k[\Phi] =
  \frac{1}{2}\mathsf{Tr}\left\{\left( \Gamma_k^{(2)}[\Phi]+R_{k}
    \right)^{-1} \partial_t R_{k}\right\}\ ,&&
\end{eqnarray}
where $t=\ln k$ denotes the logarithm of the RG scale $k$.  The trace
involves an integration over momenta or coordinates, as well as a
summation over internal spaces such as Dirac, color and flavor
indices.  $\Gamma_k^{(2)}$ represents the second functional derivative
of $\Gamma_k$ with respect to the fields $\Phi$ which together with
the regulator $R_k$ defines the inverse average propagator in
\Eq{eq:WetterichRG}.  Because $R_k$ serves as a scale dependent
infrared mass for momenta smaller than $k$ slow modes decouple from
the further evolution while high momenta are not affected. Hence,
$\Gamma_k$ interpolates between the microscopic theory at large
momenta and the macroscopic physics in the infrared (IR), $k\to0$,
where the full effective action $\Gamma \equiv \Gamma_{k\to0}$
represents the generating functional of one-particle irreducible
diagrams including all quantum fluctuations. Note that the appearance
of the full propagator turns the one-loop structure of
\Eq{eq:WetterichRG} into an exact identity and thus includes
non-perturbative effects as well as arbitrarily high loop orders. 

Without any truncations the evolved flow equations are independent of
the renormalization scheme, i.e., of the choice of the regulator
function $R_k$. However, the solution of the functional equation
requires some truncations which result in a regulator dependency.
This truncation induced dependence of physical observables on the
regulator can be minimized by choosing optimized regulators 
\cite{Litim:2001up, *Litim:2006ag}. In this
work, a modified three-dimensional version of the optimized regulator
by Litim \cite{Litim:2001up, *Litim:2006ag} is employed. For bosonic
fields the optimized regulator reads
\begin{eqnarray}
  \label{eq:reg}
  R_{k,B}(\vec p\, ) & = & \vec p^{\:2}\left( \frac{k^2}{\vec
      p^{\:2}}-1\right)\Theta\!\left(1-\frac{\vec
      p^{\:2}}{k^2}\right)\ , 
\end{eqnarray}
 and for fermions
\begin{eqnarray}
  R_{k,F}(\vec p) & = &  i\slashed{\vec p}\left(\sqrt{\frac{k^2}{\vec p^{\:2}}}-1\right)
  \Theta\!\left(1-\frac{\vec p^{\:2}}{k^2}\right) \ .
\end{eqnarray}
This choice is particularly convenient for finite temperature
calculations since the optimized flows at finite temperature factorize
\cite{Litim:2001up, Blaizot:2006rj} and the arising summation over the
Matsubara frequencies in the flow equations can be carried out
analytically.

The flow equation for $\Gamma_k$ must be supplemented with an initial
condition $\Gamma_{k\to\Lambda}$ corresponding to the microscopic
theory that is in principle given by QCD at some high initial scale
$\Lambda$. Here, we chose an initial scale of the order of $\Lambda
\approx 1$ GeV close to the threshold scale where the original QCD
degrees of freedom can be substituted by effective degrees of freedom,
cf. e.g. \cite{Braun:2009si}. This is implemented by the following
three flavor quark-meson model truncation
\begin{eqnarray}\label{eq:initaction}
 \Gamma_{\Lambda} = \!\!& 
  \displaystyle\int \!d^4 x & \bar{q} \left(\dslash  + \mu\gamma^4 +h
    \Sigma_5 \right)q\\
   && + \ \mathsf{tr}\!\left[ \partial^\mu \Sigma \partial^\mu
    \Sigma^\dag \right]
  +U_{\Lambda}\left(\rho_1,\tilde{\rho}_2,\xi\right)\ . \nonumber
\end{eqnarray}
The meson multiplets $\Sigma$ and the fields $\Sigma_5$ are given by
\Eqs{eq:sigma} and (\ref{eq:sigma5}). In the effective potential
$U_\Lambda$ a modified chiral invariant $\tilde{\rho}_2 =
\rho_2-\frac{\rho_1^2}{3}$ has been defined which simplifies some
expressions.  In the non-strange--strange ($x - y$) basis the chiral
invariants are explicitly given by\\
\begin{eqnarray}
 \rho_1 & = & \frac{1}{2}\left(\sigma_x^2+\sigma_y^2\right)\ ,\nonumber\\
  \tilde\rho_2 & = & \frac{1}{24}\left(\sigma_x^2-2\sigma_y^2\right)^2\ ,\\[1mm]
  \xi & = & \frac{\sigma_x^2\sigma_y}{2 \sqrt{2}}\ .\nonumber\\\nonumber
\end{eqnarray}
This ansatz for the effective action, \Eq{eq:initaction}, corresponds
to a derivative expansion at leading-order with a standard kinetic
term for the meson fields. No scale-dependence in the scalar
wave-function renormalizations and Yukawa coupling between the quarks
and mesons is taken into account. Finally, this yields the following
dimensionful flow equation for the grand potential
\begin{widetext}
  \begin{equation}\label{eq:flow}
    \partial_t U_k(T,\mu;\Sigma) =
    \frac{k^5}{12\pi^2}
    \left[\sum\limits_{b=1}^{2N_f^2} \frac{ 1}{E_b}\coth\left(
        \frac{ E_b}{2T}\right)  - 2N_c \sum\limits_{f=1}^{N_f} \frac{1}{E_f} 
\Biggl\{ \tanh\left( \frac{E_f + \mu}{2T}\right) + \tanh\left( \frac{E_f - \mu}{2T}\right)    \Biggr\} 
\right]\ ,
  \end{equation}
\end{widetext}
with the bosonic ($b$) and fermionic ($f$) quasi-particle energies
 \begin{equation}
 E_i = \sqrt{k^2 + m_i^2}\ ; \qquad i=b,f\ .
\end{equation} 
The masses for the quarks simplify in the ($x- y$) basis according to
\Eq{eq:mq} and the equations for the meson masses are collected in
\App{app:mesonmasses}.  In contrast to the two flavor case
\cite{Jungnickel1996b, Berges:1997eu, *Schaefer1999,
  *Braun:2003ii, Schaefer:2004en}, the isospin
symmetric potential $U_k$ now depends on two condensates, $\sigma_x$
and $\sigma_y$, denoted by $\Sigma$ in \Eq{eq:flow}, in analogy to the
mean-field potential in \Eq{eq:grandpot}.

The right-hand side of the flow equation is composed of a sum of
temperature-dependent threshold functions for each degree of freedom.
Due to the $SU(2)$ isospin symmetry some masses of the meson
multiplets degenerate and hence yield the same contribution in the
flow equation (\ref{eq:flow}).

\begin{figure*}[t!]
  \centering
  \subfigure[$\ $with  $\ua$-symmetry breaking]{\includegraphics[width=8.6cm]{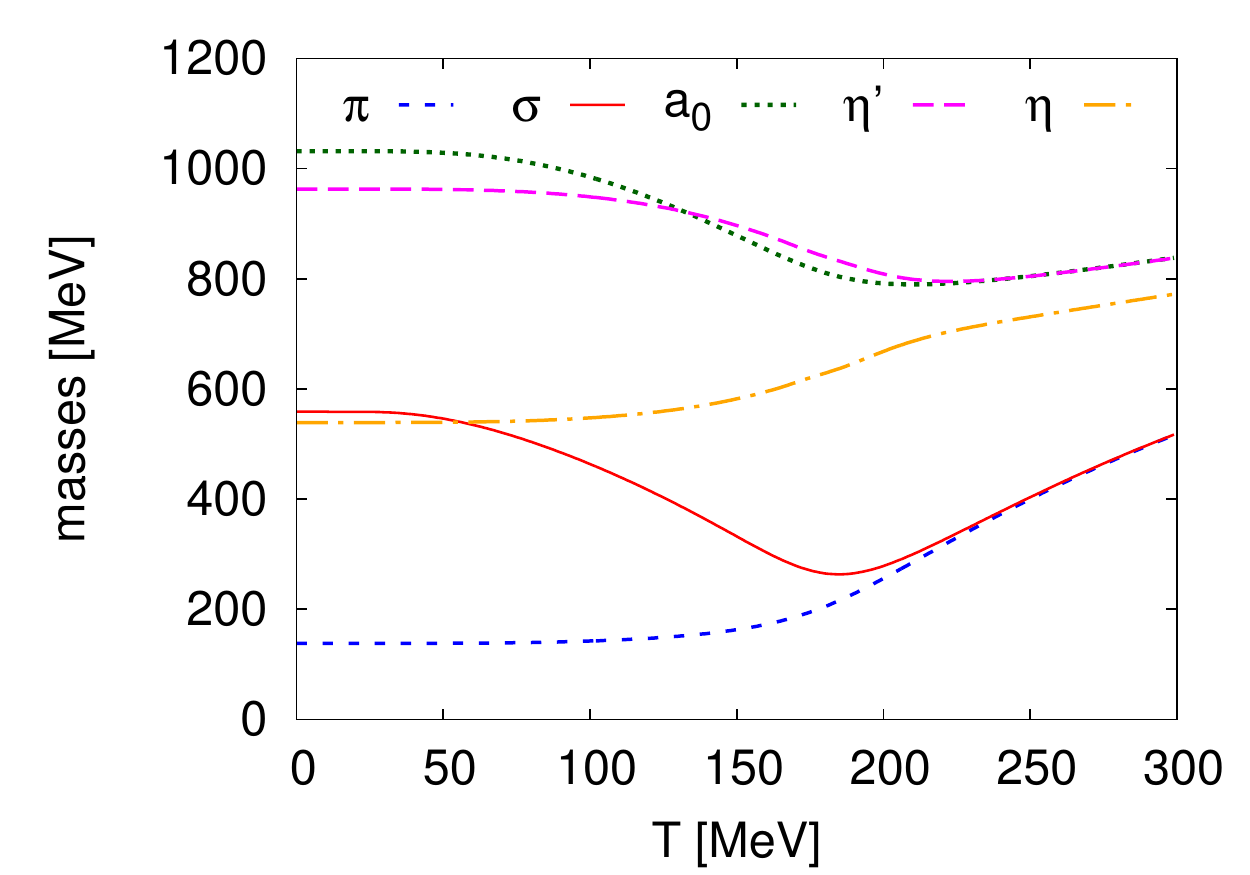}}
  \hfill
  \subfigure[$\ $without $\ua$-symmetry breaking]{\includegraphics[width=8.6cm]{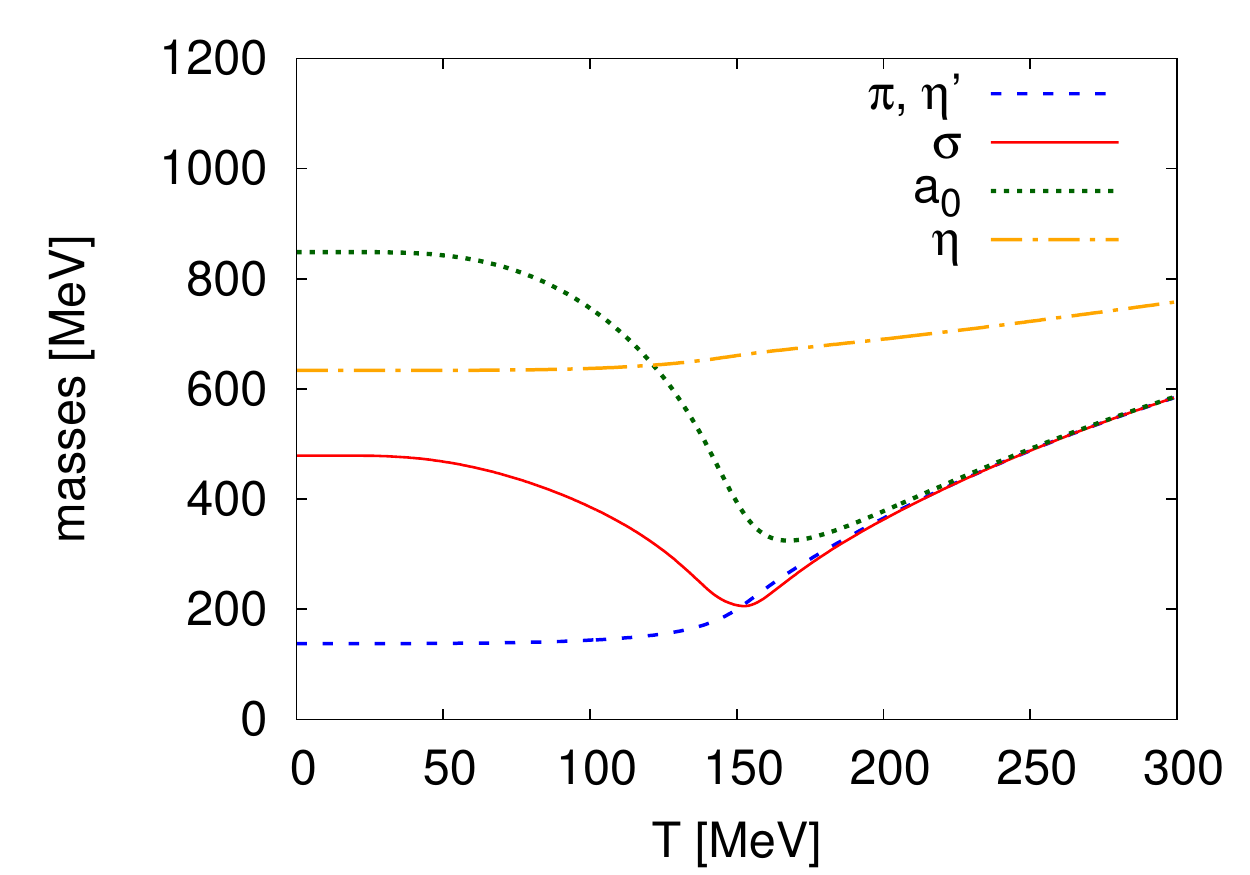}}  
  \caption{\label{fig:frg_mf_meson_masses_one_phys_mu0} Meson masses
    obtained with the FRG as a function of the temperature for
    vanishing chemical potentials.}
\end{figure*}

\section{Results and Discussion}
\label{sec:numres}

In this section we discuss and compare the phase structure of the
$(2+1)$-flavor quark-meson model in different approximations.  We
focus on the chiral phase transition at finite temperature and flavor
symmetric chemical potential and consider homogeneous chiral
condensates. Any inhomogeneities which might be of relevance at low 
temperatures and intermediate quark densities, see e.g. \cite{Nickel:2009wj}, are
thus excluded.  We thoroughly investigate the interplay of quantum and
thermal fluctuations with the anomalous $U(1)_A$-symmetry breaking via
\Eq{eq:xi}.  Consequences for the location of the critical endpoint in
the phase diagram as well as the order of the chiral transition in the
limit of vanishing light quark masses are discussed. We compare
results obtained with a full functional renormalization group (FRG)
analysis, a mean-field approximation (MFA) and with a modified FRG
approximation where meson loop contributions are dropped.  The latter
approximation is equivalent to the so-called extended mean-field
approximation where the renormalized fermionic vacuum sea term is
included in the grand potential (eMFA), see \cite{Andersen:2011pr,
  Skokov:2010sf} and Sec.~\ref{sec:mf}.  Further technical details
concerning the numerical solution of the flow equation can be found in
the appendices \ref{app:numimpl} and \ref{app:mesonmasses}. All
parameters of the model are fixed in such a way that experimental
observables like the pion decay constant and meson masses are
reproduced in the vacuum, see appendix \ref{app:numimpl_init} for
details.

\begin{figure*}[th]
  \centering 
  \subfigure[$\ $with  $\ua$-symmetry breaking]{\includegraphics[width=8.6cm]{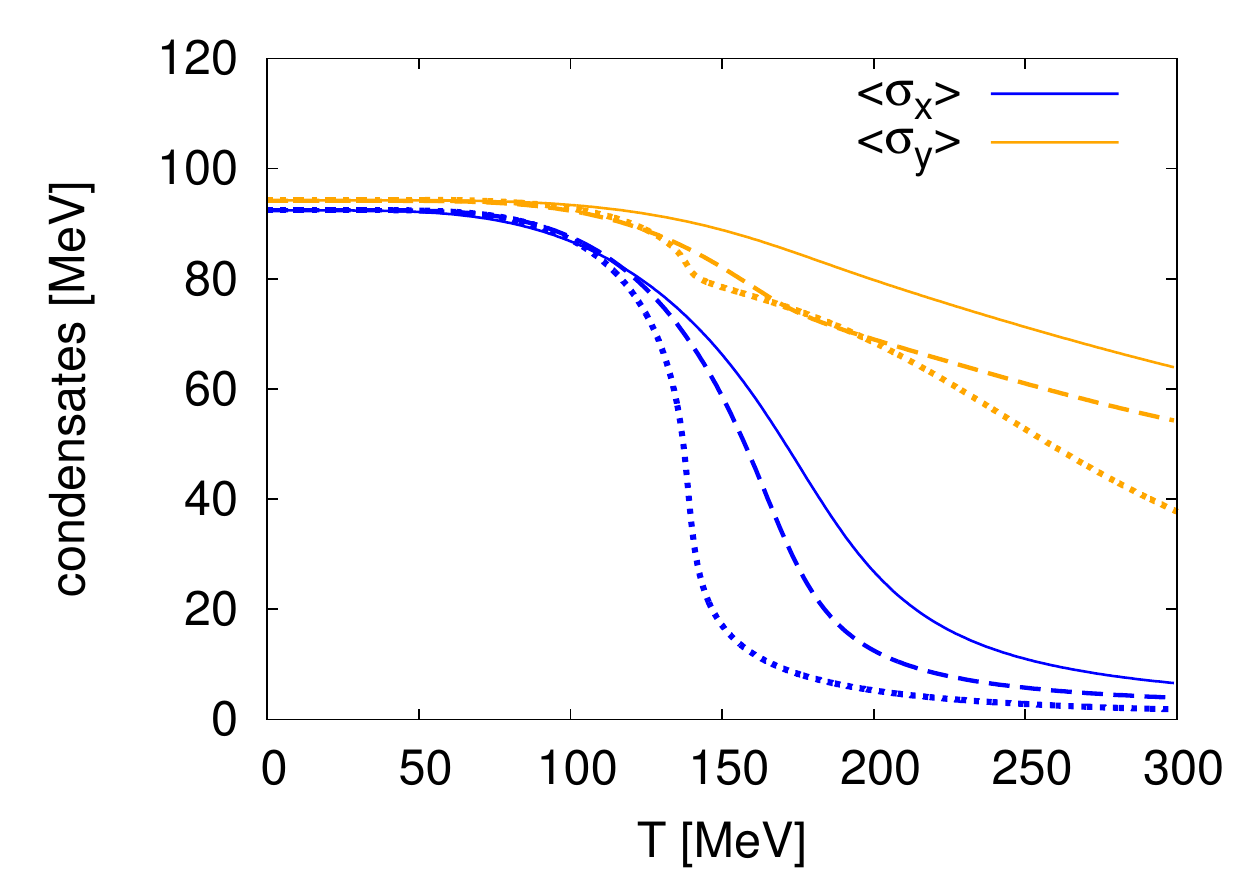}} 
  \hfill
  \subfigure[$\ $without  $\ua$-symmetry breaking]{\includegraphics[width=8.6cm]{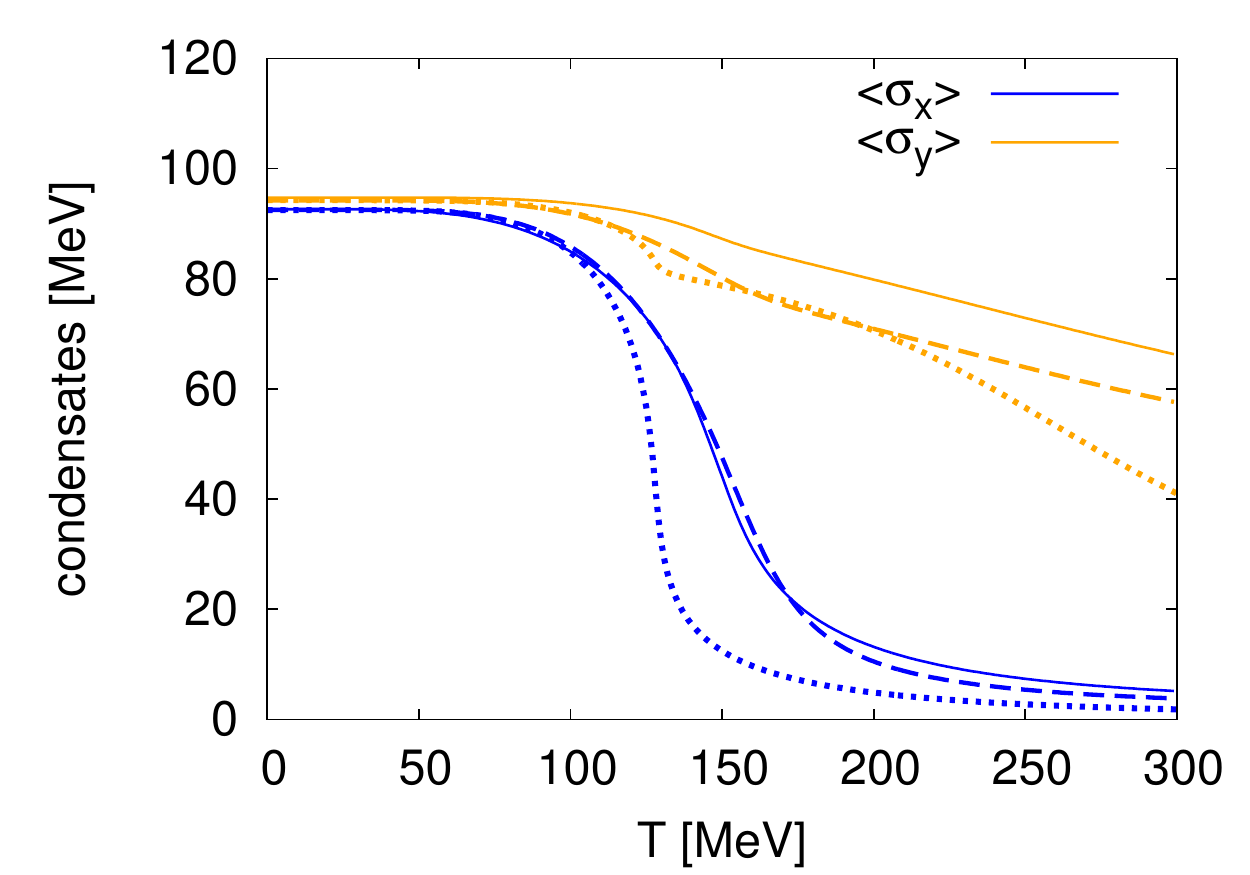}} 
  \caption{\label{fig:frg_condensates_phys_mu0} Non-strange
    $\vev{\sigma_x}$ and strange $\vev{\sigma_y}$ condensates for
    vanishing chemical potentials in different approximations (solid
    lines: FRG, dotted lines: standard MFA and dashed lines: extended
    MFA) .}
\end{figure*}

\subsection{Chiral Crossover and Axial Anomaly}

First we investigate the $U(1)_A$-symmetry breaking and the influence
of thermal fluctuations on the chiral crossover at vanishing chemical
potential $\mu=0$. In Fig.~\ref{fig:frg_mf_meson_masses_one_phys_mu0}
we show FRG results for the (pseudo)scalar non-strange and $\eta$,
$\eta'$-meson masses as a function of the temperature $T$.  In the
left panel the $\ua$-symmetry is broken via \Eq{eq:xi}. Without this
$\ua$-symmetry breaking (right panel) the $\eta'$-meson mass
degenerates always with the pion mass and the two sets of light chiral
partners $(\sigma, \vec \pi)$ and $(\vec a_0,\eta')$ collapse in the
chirally symmetric phase.

In agreement with experiment \cite{Vertesi:2009wf, *Csorgo2010},
we find for a broken $\ua$-symmetry that the mass of the $\eta'$-meson
drops around the chiral crossover.  The drop of the mass at the chiral
transition is a consequence of the Kobayashi-Maskawa-'t~Hooft term, \Eq{eq:xi}, for
three flavors.  It is cubic in the fields $\sigma_x$, $\sigma_y$ and
hence the anomalous contribution to $m^2_{\eta'}$ depends linearly on
the condensates $\vev{\sigma_x}$ and $\vev{\sigma_y}$ which both melt
at the crossover. This is demonstrated in
\Fig{fig:frg_condensates_phys_mu0}, where both condensates, including
the anomaly (left panel) and without the anomaly (right panel), are
shown as a function of the temperature for vanishing chemical
potential.  In the standard MFA the condensates decrease faster than
in approximations including fluctuations. The value of the
pseudocritical temperature, defined by the inflection point of the
non-strange condensates, varies by 10\% without the anomaly and by
20\% including the anomaly. The largest shift of the critical
temperature in comparison to the standard MFA is seen with the full
FRG: fluctuations smoothen and push the non-strange transition to higher
temperatures. This trend is
similar to previous two-flavor investigations, see
e.g.~\cite{Nakano:2009ps}.

Furthermore, the slope of the crossover is modified by mesonic
fluctuations and depends on the axial $\ua$-anomaly. Interestingly,
without the anomaly (right panel) the extended MFA and FRG results for
the non-strange condensate almost coincide for all temperatures.  With
determinant, on the other hand, the crossover is washed out even
further by the mesonic fluctuations.

The strange sector is only mildly affected by the Kobayashi-Maskawa-'t~Hooft determinant
and the strange condensate melts only moderately, see
Fig.~\ref{fig:frg_condensates_phys_mu0}.  Independent of the used
approximation the rather slow melting of the strange condensate might
be related, at least partially, to the fit procedure of the model
parameters. It has been observed previously that for low values of the
sigma meson mass, all condensates vanish in the $SU(3)$ chiral limit.
As a consequence, spontaneous chiral symmetry breaking is lost
\cite{Schaefer:2008hk} and the value of the strange condensate is mainly
governed by the large and temperature-independent explicit symmetry
breaking parameter $c_y$.  Correspondingly, we expect a considerably
smaller temperature dependence of the strange sector compared to the
light one.

Finally, the strange condensate melts even slower when mesonic
fluctuations are taken into account.  This might lead to the
conclusion that the temperature-dependence of a $\ua$-symmetry
breaking term is mostly driven by the non-strange sector. Work in this
direction is in progress \cite{mmanomalie}.

\subsection{Finite density and the critical endpoint}

In the following we explore how the location of the critical endpoint
(CEP) in the phase diagram is affected by fluctuations with and
without anomalous $\ua$-symmetry breaking.  
The location of the CEP
also depends considerably on the chosen value of the $\sigma$-meson
mass \cite{Schaefer:2008hk}.  To eliminate this $m_\sigma$-dependence,
we fix the value to $m_\sigma= 480$ MeV, unless stated otherwise.

\begin{table}[t!]
  \begin{tabular}{|c|c|c|c|}
\hline
	  $c$ [MeV] & MFA & eMFA  & FRG \\\hline
	  $4807.84$   & $(98,155)$ & $(35,293)$ & $(17,295)$\\
	  $0$         & $(93,173)$ & $(31,298)$ & $(43,280)$\\\hline
  \end{tabular}
  \caption{Critical endpoint coordinates $(T_c,\mu_c)$ in units of MeV
    for different approximations with and without anomaly.} 
  \label{tab:CEP}
\end{table}

Results for the location of the CEP obtained with our various
approximations to the grand potential are summarized in
Tab.~\ref{tab:CEP}.  In agreement with previous investigations
\cite{Schaefer:2008hk}, we find that in a standard mean-field
approximation the CEP is pushed towards smaller chemical potentials
and higher temperatures if the $\ua$-symmetry violation is taken into
account. Adding the fermionic vacuum contribution to the potential,
this behavior with respect to the anomaly does not change
qualitatively, although the CEP is pushed towards considerably larger
densities and smaller temperatures.  In contrast, by additionally
taking mesonic fluctuations into account, the dependency on the
anomalous $\ua$-symmetry breaking is reversed.  With a constant
Kobayashi-Maskawa-'t~Hooft determinant, the endpoint is pushed towards larger chemical
potentials and smaller temperatures.  It is remarkable that the
endpoint without anomaly, but with mesonic fluctuations is located at
larger temperatures and smaller chemical potentials than for the eMFA.

\begin{figure*}[t!]
  \centering \subfigure[$\ $with $\ua$-symmetry breaking]{\includegraphics[width=8.6cm]{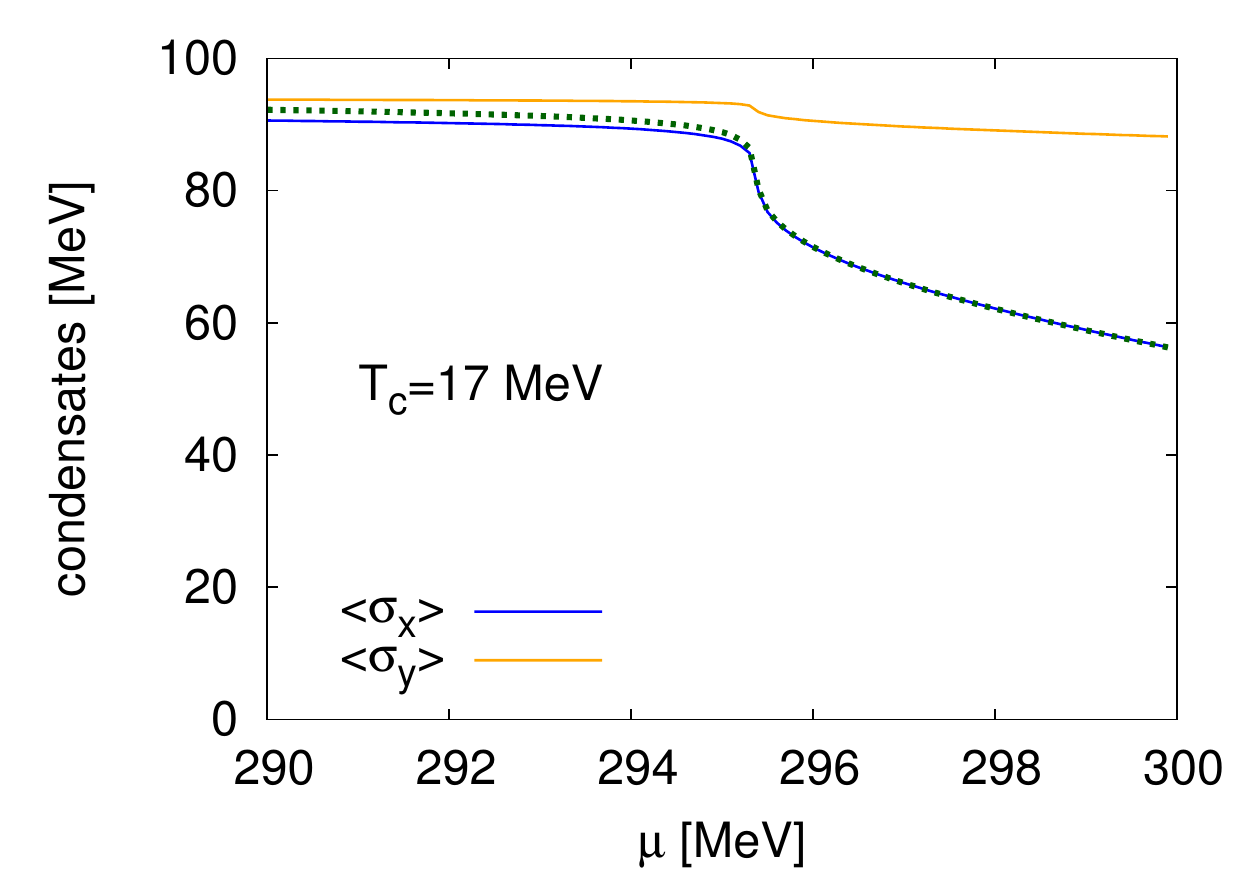}}
  \hfill
  \subfigure[$\ $without $\ua$-symmetry breaking]{\includegraphics[width=8.6cm]{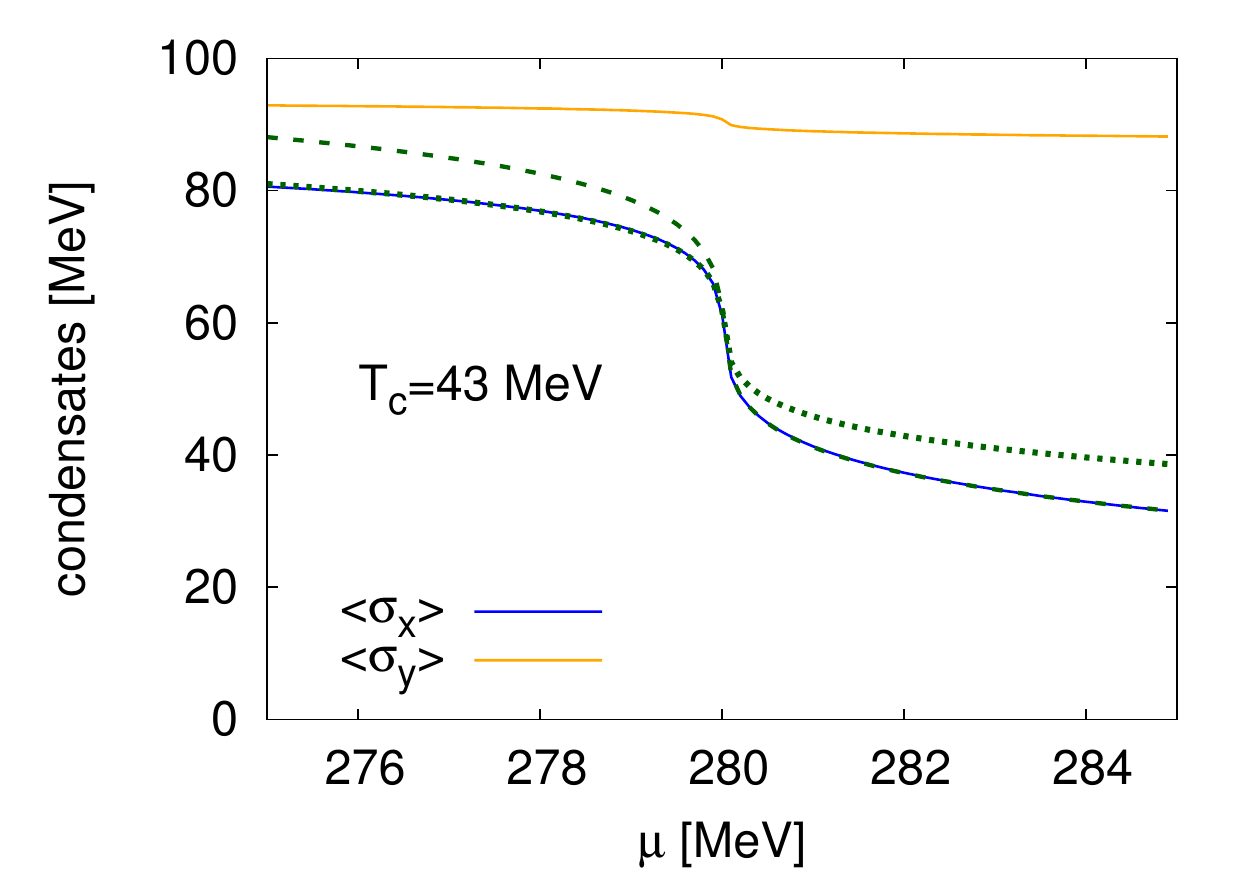}}
  \caption{\label{fig:frg_crit_mu} The condensates in the vicinity of
    the critical point as a function of the chemical potential
    ($m_\sigma=480$ MeV). Dotted and dashed lines are fit functions. See text for details.}
\end{figure*}

Although it is well-established that $\ua$-symmetry breaking terms are
suppressed at asymptotically large temperatures and chemical
potentials \cite{Gross:1980br, *Schafer:1996wv, *Schafer:2002ty}, it
is not fully settled whether this also holds at intermediate
temperatures and chemical potentials \cite{Costa:2004db}. Recent
lattice investigations indicate, however, that the $\ua$-symmetry
might already be effectively restored at temperatures slightly above
the chiral crossover \cite{Bazavov:2012qja, Cossu:2012gm,
  *Cossu:2013uua}.  Then our findings also have consequences on the
location of the CEP in two-flavor investigations.  In such studies one
usually assumes maximal $\ua$-symmetry breaking by considering only
the $(\sigma,\vec\pi)$-mesons. The $(\eta,\vec{a})$-mesons decouple
due to their assumed large masses, which are induced by the large
Kobayashi-Maskawa-'t~Hooft determinant \cite{Jungnickel1996b}.  If, however, the
$\ua$-symmetry is restored with chiral symmetry, we expect
that the critical endpoint is shifted towards larger temperatures and
lower chemical potentials also in a two flavor scenario. Thus, a
direct confirmation of this scenario requires investigations which
also include the $(\eta,\vec{a})$-mesons \cite{mmanomalie}.  This also
concerns phase structure studies in terms of QCD degrees of freedom,
e.g. \cite{Braun:2009gm, *Pawlowski:2010ht, *Fischer:2011mz,
  *Fischer:2012vc,*Fischer:2013eca}.

In \Fig{fig:frg_crit_mu} both condensates with (left) and without
anomaly (right) are plotted in the vicinity of the critical point as a
function of the chemical potential. Similar to the crossover at
$\mu=0$, the strange condensate melts considerably less than the light
condensate around the CEP in all approximations.  Interestingly, the
two condensates seem to be related to each other at criticality. We
found a scaling between both condensates that is demonstrated in
\Fig{fig:frg_crit_mu} with dashed and dotted lines which are obtained
by the ansatz $\vev \sigx = \alpha_1+\alpha_2\langle\sigma_y\rangle$
with the two constant fit parameters $\alpha_i$. In particular, with
$\ua$-symmetry breaking (left panel) we see an almost perfect scaling
in both regions while without anomaly (right panel) the scaling
depends on which phase we use to fit the parameters.

\begin{figure}[tb!]
  \centering
  \hfill
  \includegraphics[width=8.6cm]{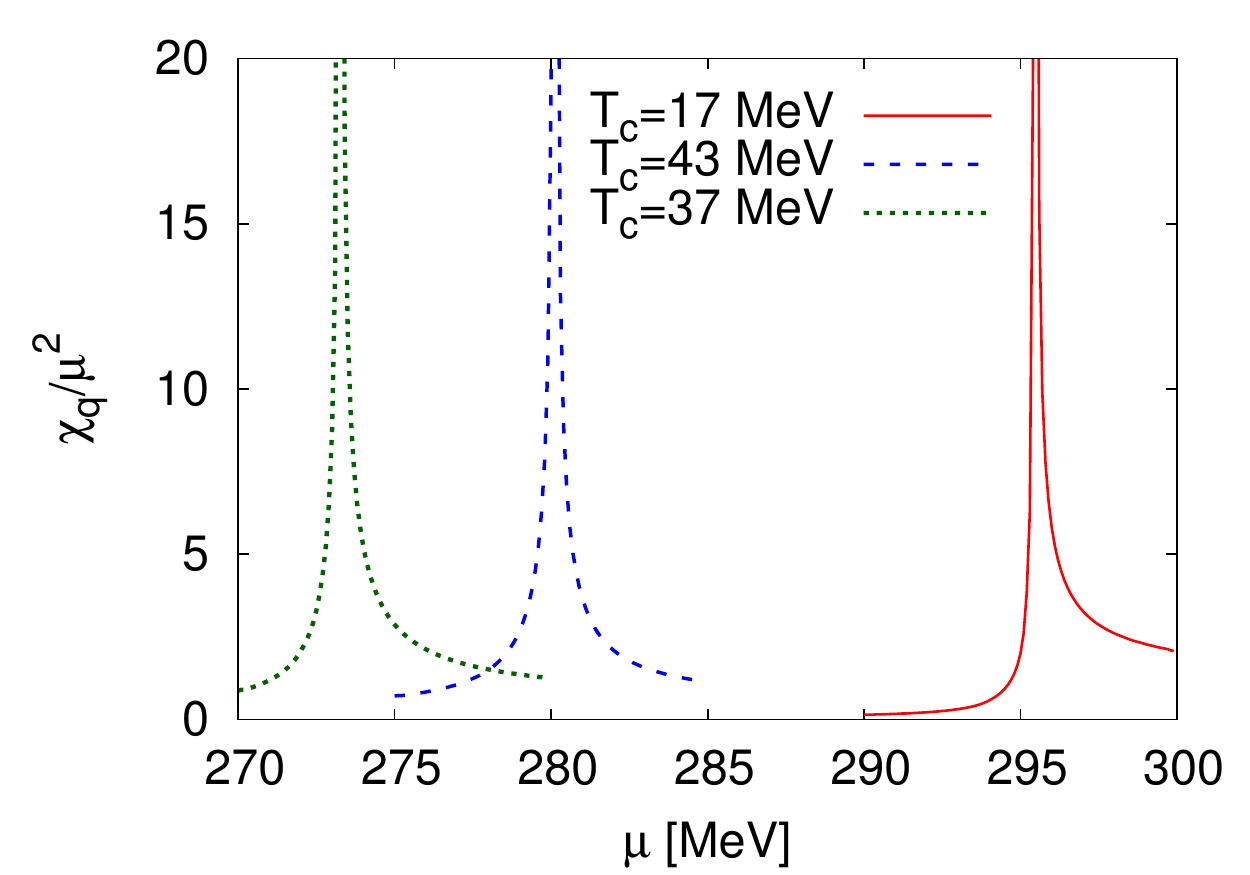}
  \caption{\label{fig:frg_crit_exp_CEP} Quark number susceptibility in
    the vicinity of the critical point without $\ua$-symmetry for
    $m_\sigma=480$ MeV (solid, red) and for $m_\sigma=400$ MeV
    (dotted, green). For comparison, the $U(1)_A$-symmetric
    susceptibility for $m_\sigma=480$ MeV (dashed, blue) is also
    shown.}
\end{figure}

The dependency of the CEP location in the phase diagram on the
$\ua$-symmetry and also on the sigma mass $m_\sigma$ is demonstrated
in \Fig{fig:frg_crit_exp_CEP}, where the FRG result for the quark
number susceptibility in the vicinity of the critical point,
normalized with the quark chemical potential, is shown as a function
of the chemical potential. In this figure two different CEP scenarios
are compared with each other.  First, fixing the sigma mass to $m_\sigma=480$
MeV (solid, red and dashed, blue lines) the point is pushed towards the
temperature axis for a $\ua$-symmetric theory.  Second, 
reducing the sigma mass to $m_\sigma =400$ MeV but fixing the
$\ua$-symmetry breaking the CEP is also pushed towards the temperature
axis (solid, red and dotted, green lines).

\begin{figure*}[ht!]
  \centering 
  \subfigure[$\ $with $\ua$-symmetry breaking]{\includegraphics[width=8.6cm]{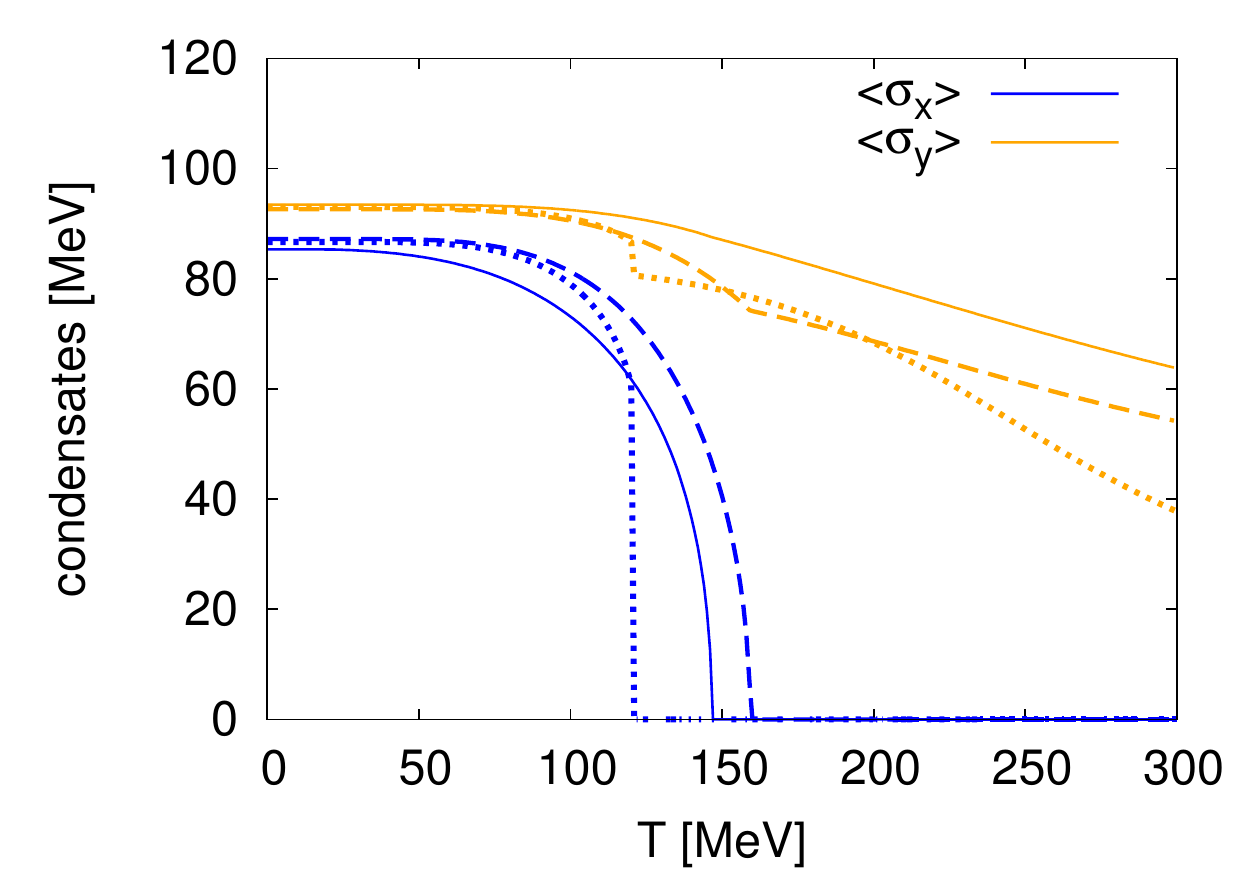}}
  \hfill
  \subfigure[$\ $without  $\ua$-symmetry breaking]{\includegraphics[width=8.6cm]{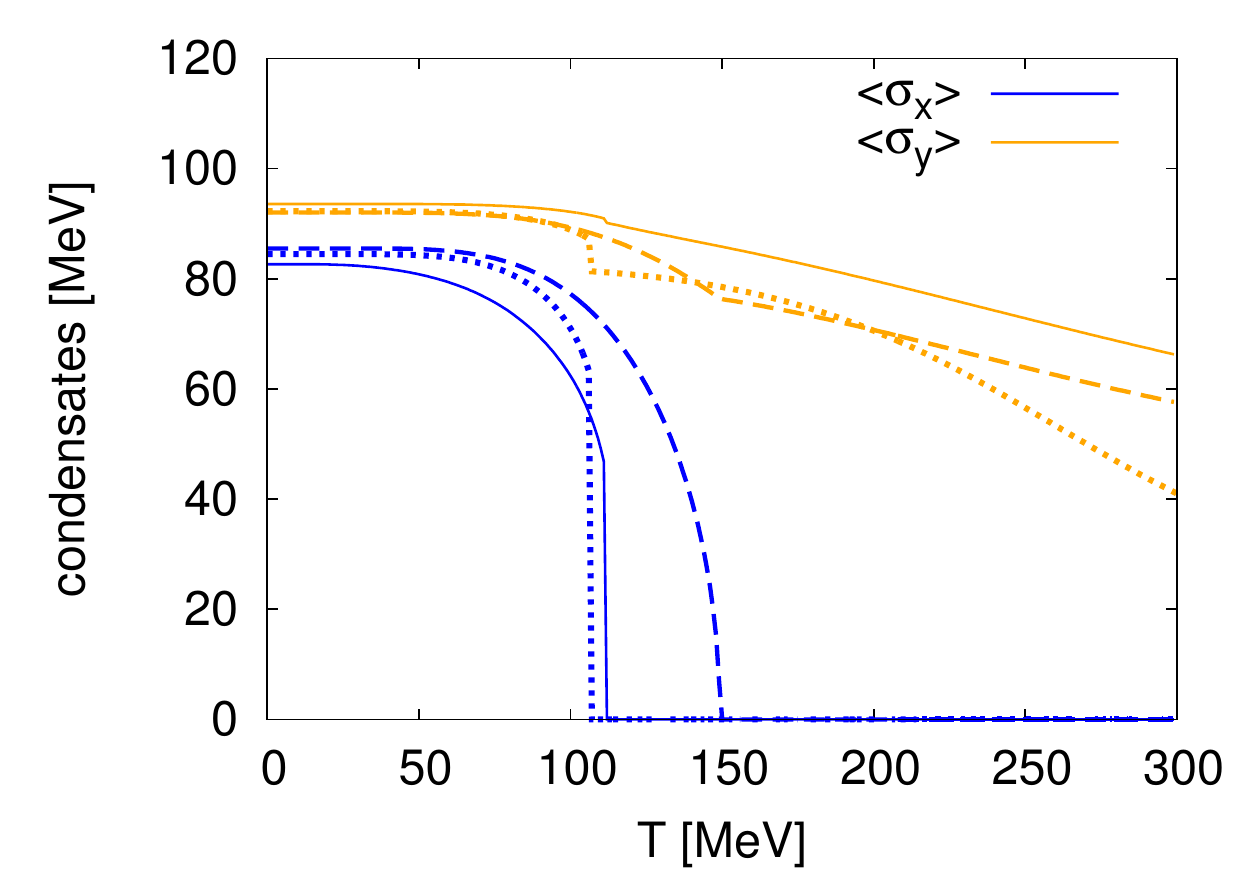}} 
  \caption{\label{fig:frg_mf_condensates_chilim_mu0} Non-strange
    $\vev{\sigx}$ and strange $\vev{\sigy}$ condensates in the
    non-strange chiral limit for $\mu=0$ obtained in different
    approximations similar to \Fig{fig:frg_condensates_phys_mu0}
    (solid lines: FRG, dotted lines: standard MFA and dashed lines:
    extended MFA).}
\end{figure*}

\subsection{Chiral Limits}

Finally, we investigate the quark mass sensitivity of the chiral
transition including the anomaly and examine various chiral limits.

Based on RG arguments for a purely bosonic theory, the chiral
transition is of first order in the $SU(3)$-symmetric chiral limit,
whereas the order of the transition in the two-flavor chiral limit
depends on the implementation of the $\ua$-symmetry breaking
\cite{Pisarski1984a}. For a $\ua$-symmetric two flavor system with
$U(2)_L\times U(2)_R$ symmetry a first-order transition is expected in
the chiral limit, see e.g.~\cite{Fukushima:2010ji, Grahl:2013pba}.
With an axial $\ua$-symmetry breaking the order can change to second
order if the coupling strength of the Kobayashi-Maskawa-'t~Hooft determinant is only
moderately temperature dependent.  Therefore, a temperature
independent, i.e. constant, $U(1)_A$-symmetry breaking can smoothen the
transition from first to second order in the two-flavor chiral limit,
whereas the opposite happens for three flavors and the first-order
transition becomes even stronger in the corresponding chiral limit.

Relative to the anomalous $\ua$-symmetry implementation, several
scenarios in the light and strange quark mass $(m_l,m_s)$-plane can
now arise. Around the $SU(3)$-symmetric chiral limit with
$(m_l,m_s)=(0,0)$ there might be a finite region of first-order
transitions with a second-order boundary line, which terminates at a
finite value of a tricritical strange mass, $m^*_s$, from which a
second-order transition line is extended to the two flavor chiral
limit. The precise location and even the existence of such a
tricritical strange quark mass $m^*_s$ is not yet fully
settled. However, there could also be a first-order region connecting
the $SU(3)$ and the two-flavor chiral limits along the $(m_l=0)$-line,
see e.g. \cite{Schaefer:2008hk, Aoki:2012yj}. Then, there should not
be a critical value of the strange quark mass $m_s^*$, but a critical
light quark mass $m_l^*$ in the two flavor case.

Independent of the $\ua$-symmetry breaking, a first-order transition
is seen in quark-meson models in standard mean-field approximations in
the two-flavor as well as in the $SU(3)$-symmetric chiral limit, see
e.g.~\cite{Schaefer:2004en, Schaefer:2008hk}.  For an explicit
symmetry breaking strength $c_y= c_{y,\text{phys}}$ that leads to a
strange constituent quark mass of $m_s = 430$ MeV we also find a
first-order transition in the non-strange chiral limit, see
\Fig{fig:frg_mf_condensates_chilim_mu0}, where both order parameters
in different approximations are plotted in the non-strange chiral
limit.  However, as already argued in \cite{Skokov:2010sf}, the
first-order transition might be misleading and an artifact of the used
mean-field approximation. Going beyond mean-field approximations by
taking the vacuum fluctuations of the quarks into account (eMFA), this
behavior is changed and a second-order transition is observed, which
is independent of the axial anomaly (see dashed lines in the
figure). When mesonic fluctuations are considered in addition via a
full FRG treatment, the order of the transition depends on the
anomaly. Including the Kobayashi-Maskawa-'t~Hooft determinant, a
second-order transition is found while for a $\ua$-symmetric theory
the transition becomes first-order (solid lines).

This agrees with the picture that the chiral transition is mostly
driven by two flavor dynamics, i.e., the non-strange chiral limit
behaves qualitatively more like the $SU(2)$-symmetric than the
$SU(3)$-symmetric chiral limit.  As a consequence, the
$U(1)_A$-symmetry violating and temperature independent determinant
acts like a mass term, which leads to a second-order
transition as in \cite{Pisarski1984a}.

\begin{figure}[tb!]
  \centering
  \hfill
  \includegraphics[width=8.6cm]{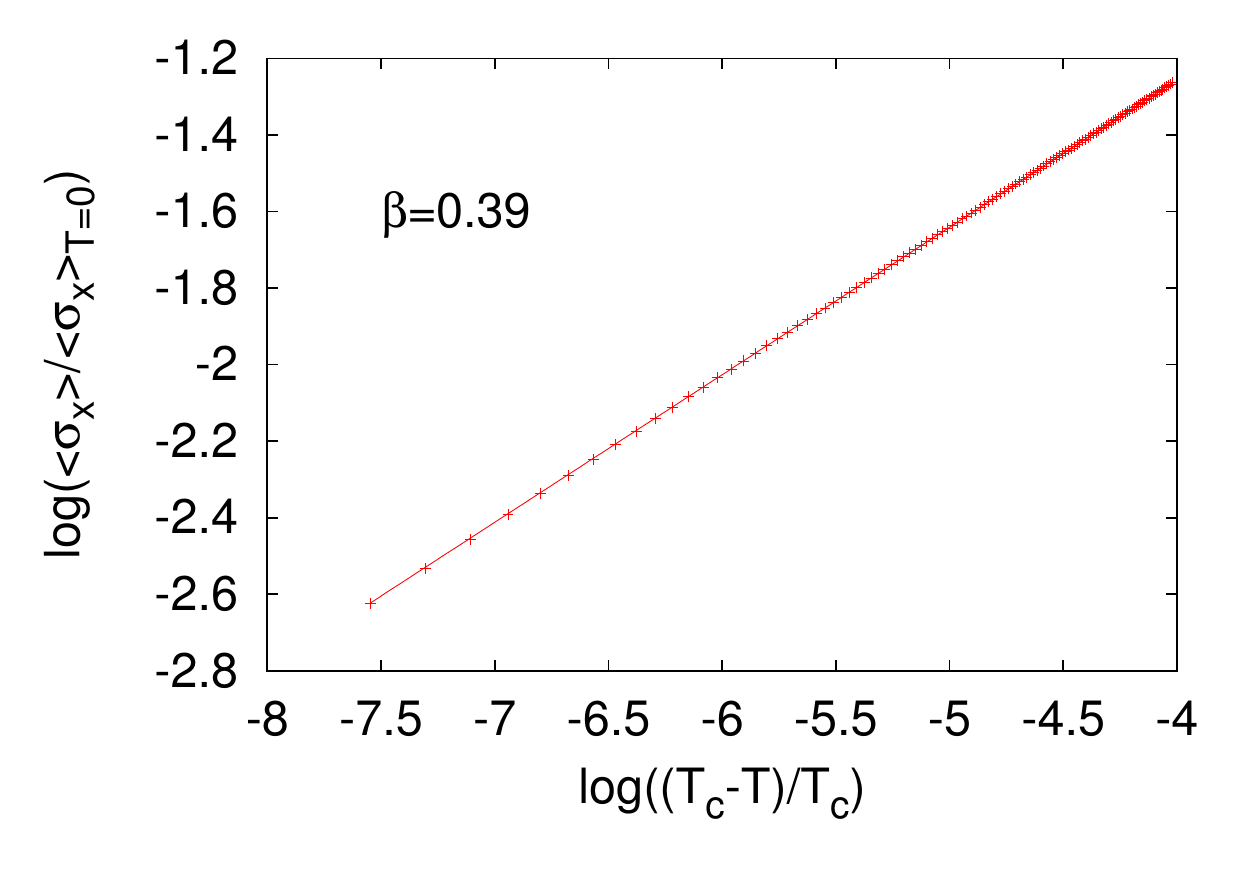}
  \caption{\label{fig:beta_exp} Scaling of the non-strange
    condensate close to the critical point. The slope corresponds to
    the $O(4)$ critical exponent $\beta=0.39$.}
\end{figure}

Without the Kobayashi-Maskawa-'t~Hooft term we have also investigated the shape of the
first-order region by varying the explicit symmetry breaking parameter
$c_y$ around its physical value $c_{y,\text{phys}}$. In this way we
have determined with the FRG the corresponding $c^*_{x}$ that leads to
a second-order chiral transition.  Since the value of the critical
$c^*_{x}$ also depends on the chosen coarse graining scale in the
infrared \cite{Litim:1994jd, *Litim:1996nw, *Berges:1996ib}, we can
only determine a qualitative picture of the first-order region. For
example, in \Fig{fig:frg_mf_condensates_chilim_mu0} an infrared cutoff
of the order of $100$ MeV has been employed.  Schematically, we find
that the critical $c^*_{x}$ grows with $c_y$ around its physical value
$c_{y,\text{phys}}$.  A similar behavior has also been found in a
purely mesonic study \cite{Lenaghan:2000kr} and is consistent with the
scenario that the first-order region is extended along the
$(m_l=0)$-line without anomalous $U(1)_A$-symmetry breaking.

For the second-order transition, we find critical exponents which lie
in the $O(4)$-universality class. For example, we find $\beta=0.39$,
as demonstrated in the \Fig{fig:beta_exp}, where the scaling of the
order parameter over several orders of magnitudes is displayed. Since
the anomalous dimension vanishes in leading-order derivative expansion
of the average effective action, $\eta \equiv 0$, the remaining
critical exponents are given by (hyper)scaling relations.  The $O(4)$
critical exponents are a consequence of the finite Kobayashi-Maskawa-'t~Hooft coupling
at the critical temperature. If, however, the $U(1)_A$-symmetry were
effectively restored at the chiral transition, we would expect
critical exponents in the $U(2)_L\times U(2)_R/U(2)_V$ universality
class in case of a second-order transition \cite{Basile:2005hw,
  *Vicari:2007ma, Aoki:2012yj, Grahl:2013pba}.

\section{Summary and conclusions}
\label{sec:summary}

We have investigated the consequences of anomalous $\ua$-symmetry
breaking in the presence of quantum and thermal fluctuations in a
three flavor effective description for QCD with a focus on the phase
structure.  With an effective quark-meson model a drop of the
anomalous mass of the $\eta'$-meson at the chiral crossover
temperature is found which agrees well with recent experiments.  In
analogy to a corresponding two flavor description fluctuations weaken
the chiral crossover and the chiral condensates decrease less rapidly.
In particular, the strange condensate melts considerably slower when
fluctuations are taken into account. This leads to the conclusion that
the chiral dynamic around the crossover is predominantly governed by
the light quark sector.

At finite quark chemical potential, mesonic fluctuations lead to a
considerable effect in the presence of an anomalous $\ua$-symmetry
breaking realized by a Kobayashi-Maskawa-'t~Hooft determinant in the Lagrangian. Without
anomalous $\ua$-symmetry breaking the endpoint is pushed to
significantly larger temperatures and smaller chemical potentials. This
strong dependency of the CEP location on $\ua$-symmetry breaking is
opposite to what is found in corresponding mean-field investigations.
Hence, for future investigations on the existence/location
of a possible critical endpoint in the QCD phase diagram it is crucial
to consider all quantum and thermal fluctuations with a proper
$\ua$-symmetry breaking and its possible effective restoration.

Finally, we have investigated the order of the chiral transition in
the limit of vanishing light, but physical strange quark masses. In
standard mean-field approximations we find, independent of the
anomalous $\ua$-symmetry breaking, a first-order transition and a
second-order transition if the vacuum term is included. However, with
mesonic fluctuations the transition is of first-order without the
axial $\ua$-anomaly and of second-order with the anomaly, lying in the
$O(4)$-universality class. This demonstrates once more the importance
of the interrelation of mesonic fluctuations with the chiral
$\ua$-anomaly.  Furthermore, in agreement with other model studies
without quarks, e.g.~\cite{Lenaghan:2000kr}, we find for growing light
quark masses that the first-order region in the ($m_l$, $m_s$)-plane
is extended to larger strange quark masses if the anomaly is
neglected.

\subsection*{Acknowledgments}

We thank T.K. Herbst, J.M. Pawlowski, R. Stiele and M. Wagner for
interesting and enlightening discussions.  M.M. acknowledges support
by the FWF through DK-W1203-N16, the Helmholtz Alliance HA216/EMMI,
and the BMBF grant OSPL2VHCTG. B.-J.S. acknowledges support by the FWF
grant P24780-N27 and by CompStar, a research networking programme of
the European Science Foundation.  This work is supported by the
Helmholtz International Center for FAIR within the LOEWE program of
the State of Hessen.

\appendix

\section{Numerical implementations}
\label{app:numimpl}

In this appendix some technical details of the used two-dimensional
grid and Taylor expansion technique for the numerical solution of the
flow equation are given. At the end of this appendix the procedure for
fixing the initial parameters for the flow equations is summarized.

\subsection{Two-dimensional grid and Taylor technique}

For the construction of the two-dimensional grid we define the
positive variables
\begin{eqnarray}
 x = \sigma_x^2,\quad y = 2\sigma_y^2-\sigma_x^2\ .
\end{eqnarray}
Clamped cubic splines are used to evaluate the required first- and
second-order derivatives of the effective potential as a function of
$x$ and $y$, see also \cite{Strodthoff:2011tz}. Derivatives of
the effective potential
with respect to the chiral
invariants are obtained by applying the chain rule.

For comparison, a Taylor expansion of the effective potential in the
variables $\rho_1$ and $\tilde\rho_2$ is performed to order
$\mathcal{O}\left(\rho_i^3\right)$ around the $k$-dependent minimum.
The scale dependence of the minimum is used to replace the flow of the
Taylor coefficients $a_{10}$ and $a_{01}$ via the conditions
\begin{eqnarray}
 \partial_{\sigma_x} U_k\big\vert_{\text{min}} =  \partial_{\sigma_y} U_k\big\vert_{\text{min}} = 0 \ .
\end{eqnarray}
The convergence properties of such expansion schemes have
been studied in, e.g., \cite{Tetradis:1993ts,
  Papp2000}.

\begin{figure}[t!]
  \centering
  \hfill
  \includegraphics[width=8.6cm]{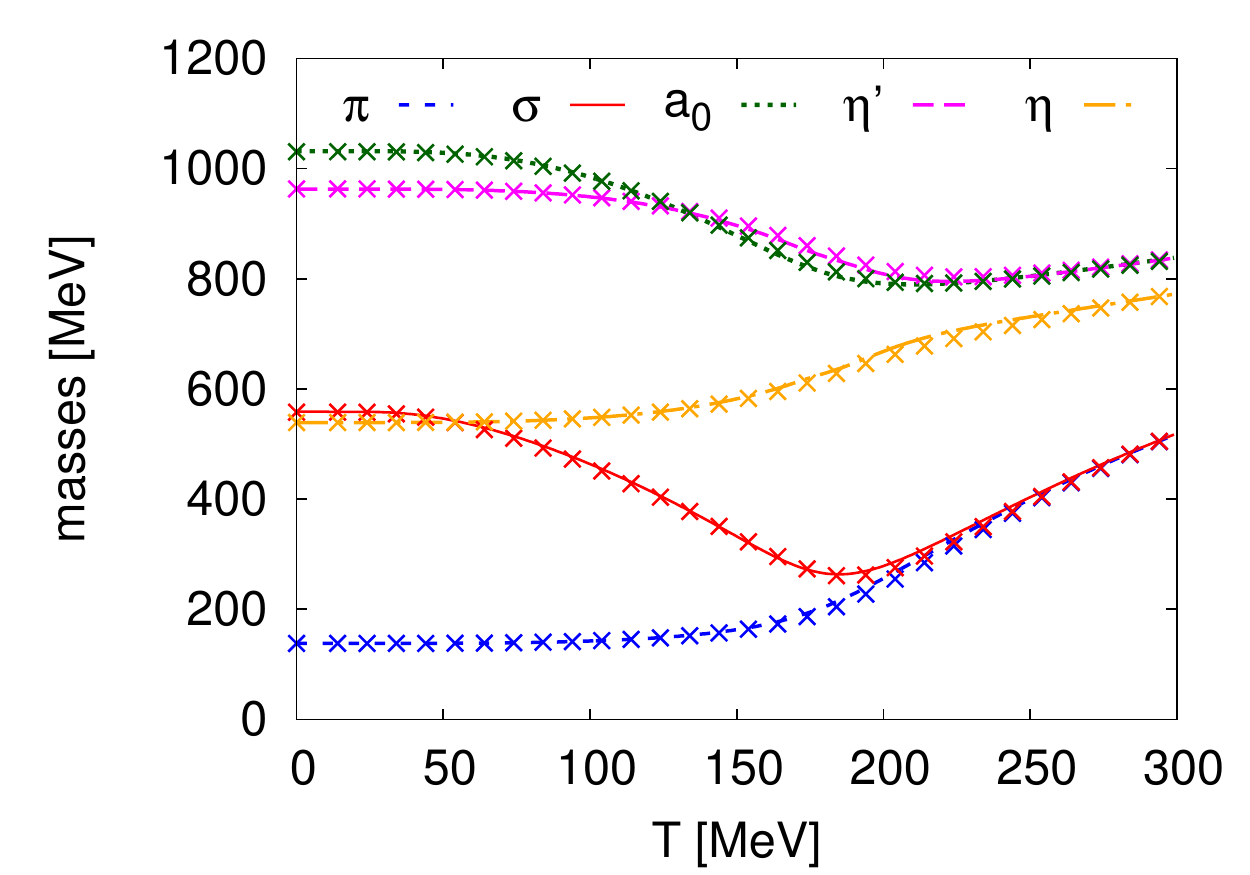}
  \caption{\label{fig:frg_crit_mu3} Meson masses obtained
    with the grid method (solid lines) and with the Taylor expansion
    technique (crosses).}
\end{figure}

In both numerical approaches, the right hand side of the flow equation
(\ref{eq:flow}) requires the knowledge of the eigenvalues of the
Hessian of the effective potential with respect to all mesonic fields
$\Sigma$. The Hessian is given in terms of derivatives of the
effective potential with respect to the chiral invariants $\rho_1$ and
$\tilde\rho_2$. The resulting expressions are summarized in
\App{app:mesonmasses}.  Then, the flow equation becomes a coupled set
of non-linear ordinary differential equations for either the values of
the effective potential at the interpolated grid points or for the
Taylor expansion coefficients (beta-functions), which can be solved
with standard numerical methods. A numerical comparison of both
techniques is given in \Fig{fig:frg_crit_mu3}, where the meson masses
are plotted as a function of the temperature for $\mu=0$. Both methods
agree very well, at least for small chemical potentials, see also
\cite{Nakano:2009ps}.

\subsection{Initial condition}
\label{app:numimpl_init}

In order to solve the flow equation the initial action has to be
specified at a given UV scale.  Explicitly, as initial potential at
$\Lambda=1$ GeV we use the expression
\begin{eqnarray}\label{eq:initpotential}
  U_{\Lambda} & = & \tilde{U}_{\Lambda}(\rho_1,\tilde{\rho}_2)- c\xi -
  c_x\sigma_x-c_y\sigma_y
\end{eqnarray}
with
\begin{eqnarray}
  \tilde{U}_{\Lambda}(\rho_1,\tilde{\rho}_2) & = &
  a_{10,\Lambda}\rho_1 + \frac{a_{20,\Lambda}}{2}\rho_1^2 +
  a_{01,\Lambda}\tilde{\rho}_2\ .\nonumber 
\end{eqnarray}
and fix the Yukawa coupling $h$ such that physical values for the pion
and kaon decay constants $f_\pi$ and $f_K$, the pion and kaon masses
$m_\pi$ and $m_K$, the combined $\eta$ and $\eta'$ masses
$m_\eta^2+m_{\eta'}^2$, the sigma mass $m_\sigma$ and the light
constituent quark mass $m_l$ are obtained in the infrared.  In
contrast to a mean-field treatment \cite{Schaefer:2008hk} it is not
possible to choose arbitrary values for the mesonic potential in the
infrared.  
Especially, the sigma meson mass is
restricted to values between approximately $400$ and $600$ MeV
for physical pion masses $m_\pi\approx140$ MeV.
We use $h=6.5$,
$c_x=(120.73$~MeV$)^3$ and $c_y=(336.41$~MeV$)^3$ together with
$a_{10,\Lambda}$, $a_{01,\Lambda}$ and $a_{20,\Lambda}$ as given in
\Tab{tab:parameter}.
\begin{table}[h!]
\begin{tabular}{|c|c|c|c|c|}
\hline
	  $m_\sigma$ [MeV] 	& $c$ [MeV] 	& $a_{10,\Lambda}$ [MeV$^2$]  	& $a_{01,\Lambda}$ 	& $a_{20,\Lambda}$ \\\hline
	  $480$ 		& $0$   	& $(178.88)^2$ 			& $140$ 		& $25$\\
	  $400$ 		& $4807.84$   & $(972.63)^2$ 			& $50$ 			& $2.5$\\
	  $480$ 		& $4807.84$   & $(867.76)^2$ 			& $50$			& $12$\\
	  $560$ 		& $4807.84$   & $(542.22)^2$ 			& $50$ 			& $36$\\\hline
  \end{tabular}
\caption{\label{tab:parameter} Parameters for the initial potential.}
\end{table}
\noindent 
Fixing the parameters in the vacuum, the finite temperature results
are then predictions.

\section{Meson Masses}
\label{app:mesonmasses}

In this section the explicit expressions for the mesonic screening
masses are collected. They are derived from the potential
\begin{eqnarray}
 U_k(\Sigma) & = & \tilde{U}_k(\rho_1,\tilde{\rho}_2)-c\xi-c_x\sigma_x-c_y\sigma_y\ ,
\end{eqnarray}
with isospin symmetry, i.e., for $\langle\sigma_3\rangle = 0$.

The squared masses are defined by the eigenvalues of the Hessian
matrix of the effective potential   $\nabla_\Sigma^2 U_k  =  \nabla_\Sigma^2 \tilde{U}_k-c
  \nabla_\Sigma^2 \xi$ with
\begin{eqnarray}
\nabla_\Sigma^2 \tilde{U}_k  & = &  \left(\partial_{\rho_1}\tilde{U}_k\right)
  \nabla_\Sigma^2\rho_1 +
  \left(\partial_{\tilde\rho_2}\tilde{U}_k\right)
  \nabla_\Sigma^2\tilde\rho_2   \\
  &&\hspace*{-2.5ex} + \left(\partial_{\rho_1}^2\tilde{U}_k\right)
  (\nabla_\Sigma\rho_1)^T \nabla_\Sigma\rho_1 -  c \nabla_\Sigma^2 \xi\nonumber \\
  &&\hspace*{-2.5ex}+ \left(\partial_{\tilde\rho_2}^2\tilde{U}_k\right) (\nabla_\Sigma\tilde\rho_2)^T \nabla_\Sigma\tilde\rho_2\nonumber\\
  &&\hspace*{-8.5ex} +\left(\partial_{\rho_1}\partial_{\tilde\rho_2}\tilde{U}_k\right)
  \left[(\nabla_\Sigma\rho_1)^T
    \nabla_\Sigma\tilde\rho_2+(\nabla_\Sigma\tilde\rho_2)^T
    \nabla_\Sigma\rho_1\right] \ .\nonumber
\end{eqnarray}
Hence, the mesonic masses can be calculated from derivatives of
$\tilde{U}_k$ with respect to the invariants $\rho_1$ and
$\tilde\rho_2$ together with the gradient and Hessian of the
invariants with respect to
$\Sigma=(\sigma_x,\sigma_1,\dots\sigma_7,\sigma_y,\pi_0,\dots,\pi_8)$.
The latter are given by

\begin{align}
 (\nabla_\Sigma\rho_1)_1  =  \sigma_x\ ,&\qquad
 (\nabla_\Sigma\rho_1)_8  =  \sigma_y\ ,\\[1ex]
 (\nabla_\Sigma^2\rho_1)_{i,i}  =  1\ ,&\quad i=1,\dots,18\ ,
\end{align}

\begin{align}
 (\nabla_\Sigma\tilde\rho_2)_1 & =  \sigma_x\frac{(\sigma_x^2-2\sigma_y^2)}{6}\ ,\\
 (\nabla_\Sigma\tilde\rho_2)_8 & =  -\sigma_y\frac{(\sigma_x^2-2\sigma_y^2)}{3}\ ,\nonumber
\end{align}

\begin{subequations}
\begin{align}
 (\nabla_\Sigma^2\tilde\rho_2)_{1,1} & =  \frac{3\ \sigma_x^2-2\ \sigma_y^2}{6}\ ,\nonumber\\
 (\nabla_\Sigma^2\tilde\rho_2)_{9,9} & =  \frac{-\sigma_x^2+6\ \sigma_y^2}{3}\ ,\\
 (\nabla_\Sigma^2\tilde\rho_2)_{1,9} & =  -\frac{2\ \sigma_x\sigma_y}{3}\ ,\nonumber
\end{align}

\begin{align}
 (\nabla_\Sigma^2\tilde\rho_2)_{i,i} & =  \frac{7\ \sigma_x^2-2\ \sigma_y^2}{6}\ ,\quad i=2,3,4\ ,\\
 (\nabla_\Sigma^2\tilde\rho_2)_{i,i} & =  \frac{\sigma_x^2+\sqrt{18}\
   \sigma_x\sigma_y+4\ \sigma_y^2}{6}\ ,\quad  i=5,\dots,8\ ,\nonumber
\end{align}

\begin{align}
 (\nabla_\Sigma^2\tilde\rho_2)_{10,10} & =  0\ ,\nonumber\\
 (\nabla_\Sigma^2\tilde\rho_2)_{18,18} & =  \frac{-\sigma_x^2+2\ \sigma_y^2}{6}\ ,\\
 (\nabla_\Sigma^2\tilde\rho_2)_{10,18} & =  \frac{\sigma_x^2-2\ \sigma_y^2}{3\sqrt{2}}\ ,\nonumber
\end{align}

\begin{align}
 (\nabla_\Sigma^2\tilde\rho_2)_{i,i} & =  \frac{\sigma_x^2-2\ \sigma_y^2}{6}\ ,\quad i=11,12,13\ ,\\
 (\nabla_\Sigma^2\tilde\rho_2)_{i,i} & =  \frac{\sigma_x^2-\sqrt{18}\ \sigma_x\sigma_y+4\ \sigma_y^2}{6}\ ,\quad i=14,\dots,17\ ,\nonumber
\end{align}
\end{subequations}

\begin{subequations}
\begin{align}
 (\nabla_\Sigma^2\xi)_{1,1} & =  \frac{\sigma_y}{\sqrt{2}}\ ,\nonumber\\
 (\nabla_\Sigma^2\xi)_{9,9} & =  0\ ,\\
 (\nabla_\Sigma^2\xi)_{1,9} & =  \frac{\sigma_x}{\sqrt{2}}\ ,\nonumber\\\nonumber
\end{align}

\begin{align}
 (\nabla_\Sigma^2\xi)_{i,i} & =  -\frac{\sigma_y}{\sqrt{2}}\ ,\quad i=2,3,4\ ,\\
 (\nabla_\Sigma^2\xi)_{i,i} & =  -\frac{\sigma_x}{2}\ ,\quad i=5,\dots,8\ ,\nonumber
\end{align}

\begin{align}
 (\nabla_\Sigma^2\xi)_{10,10} & =  \frac{-2\ \sigma_x-\sqrt{2}\ \sigma_y}{3}\ ,\nonumber\\
 (\nabla_\Sigma^2\xi)_{18,18} & =  \frac{5\ \sigma_x-\sqrt{2}\ \sigma_y}{6}\ ,\\
 (\nabla_\Sigma^2\xi)_{10,18} & =  \frac{\sqrt{2}\ \sigma_x-2\ \sigma_y}{6}\ ,\nonumber
\end{align}
\begin{align}
 (\nabla_\Sigma^2\xi)_{i,i} & =  \frac{\sigma_y}{\sqrt{2}}\ ,\quad i=11,12,13\ ,\\
 (\nabla_\Sigma^2\xi)_{i,i} & =  \frac{\sigma_x}{2}\ ,\quad i=14,\dots,17\nonumber
\end{align}
\end{subequations}

\noindent where the subscripts $i$ denote the components of the vectors and
matrices, respectively.  Hence, the Hessian splits into a
(pseudo)scalar $(9\times 9)$-mass matrix ($M^2_{ps}$) $M^2_s$
\begin{eqnarray}
\nabla_\Sigma^2 U_k & = & \begin{pmatrix}
 M^2_{s} & 0 \\
 0 & M^2_{ps} 
\end{pmatrix}\ ,
\end{eqnarray}
where the only nonvanishing off-diagonal entries in each submatrix are
the $1$, $9$ components.

Note that in a mean-field treatment there is an additional quark
contribution to the effective potential, which also modifies the
Hessian, see Ref.~\cite{Schaefer:2008hk} for explicit expressions.
For a renormalization group treatment, on the other hand, these
contributions are naturally included in the IR solution of the
potential $U_{k\rightarrow 0}$.

\end{document}